\documentclass[%
 reprint,
superscriptaddress,
noeprint,
 amsmath,amssymb,
 aps,
 prd,
floatfix,
longbibliography,
]{revtex4-2}

\usepackage{graphicx}
\usepackage{dcolumn}
\usepackage{booktabs}
\usepackage[mathlines]{lineno}

\usepackage[T1]{fontenc}
\newcommand\pt{p_{\text{T}}}
\newcommand\Et{E_{\text{T}}}
\newcommand\xt{x_{\text{T}}}
\newcommand\kt{k_{\text{T}}}
\newcommand\pythia{\textsc{Pythia}}
\newcommand\geant{\textsc{GEANT}}
\newcommand\ourabstract{Jets are collimated clusters of particles formed by the hadronization of partons following a hard interaction.  In proton-proton ($pp$) collisions at the Relativistic Heavy Ion  Collider (RHIC), jet production is dominated by $gg$ and $qg$ partonic processes, allowing us to directly probe the gluon parton distribution function (PDF) in the proton in a way complementary to deep inelastic scattering. In this paper, we report the double-differential inclusive-jet cross sections as a function of jet transverse momentum, $\pt{}$, and pseudorapidity, $\eta$, at center-of-mass energies $\sqrt{s} = 200$ and $510$~GeV, from $pp$ collisions studied with the STAR detector.  The jet $\pt$ is corrected for underlying event contributions by applying an off-axis cone method. At mid-pseudorapidity, $|\eta| < 0.9$, the kinematic coverage of our data extends to $0.07 < \xt{}\text{ (}= 2\pt{} / \sqrt{s} \text{)} < 0.5$ and $0.03 < \xt{} < 0.31$ at $\sqrt{s} = 200$~and 510 GeV, respectively, where the gluon PDF is poorly constrained by the TeV-scale $pp$~($p\bar{p}$) colliders. The inclusive jet cross sections are compared to the next-to-next-to-leading order perturbative quantum chromodynamics calculations using several recent PDF sets as inputs. These results will further constrain the gluon PDF, help tune Monte Carlo generators, and  provide critical reference data needed to study the quark-gluon plasma.}
\newcommand\figureone{%
\includegraphics{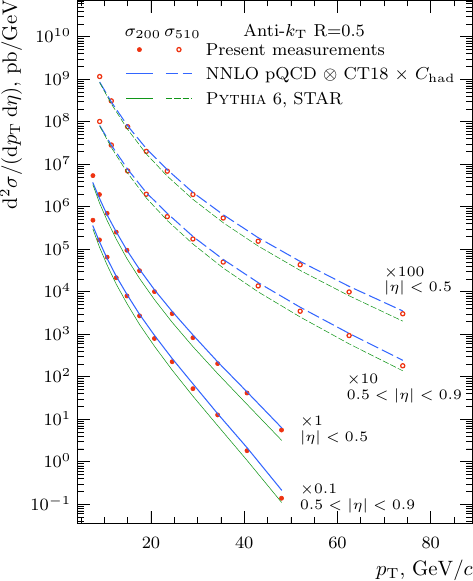}%
\caption{\label{fig:pp_xsec} Unfolded double-differential jet cross section, $\frac{d^2\sigma}{d\pt{}d\eta}$, as a function of the jet $\pt{}$ at $\sqrt{s} = 200$~and 510 GeV. The double-differential jet cross sections from the NNLO pQCD with CT18 PDF after hadronization corrections and the \pythia{} 6 STAR tune are also plotted.
}
}
\newcommand\figurethree{%
\includegraphics{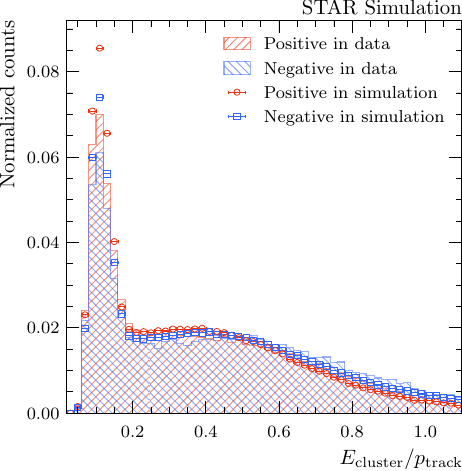}
\caption{\label{fig:ecpvspt} A comparison between data and simulation of the charge-separated $f_{h}$ 
distribution for tracks with momenta, 2 $<p<$~3 GeV/$c$ from 510 GeV $pp$ collisions. The reason to separate by hadron charge is that $K^{-}$ has a larger interaction cross section with the electromagnetic calorimeter materials than $K^{+}$, and that $\bar{p}$ can annihilate with $p$ from the $\text{Pb}$ plates in the calorimeter. 
}
}

\newcommand\figurefour{%
\includegraphics{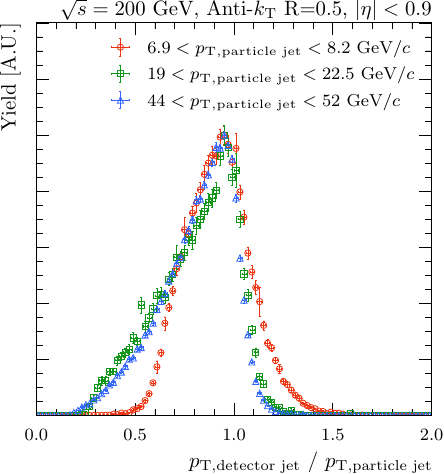}%
\hspace{20pt}%
\includegraphics{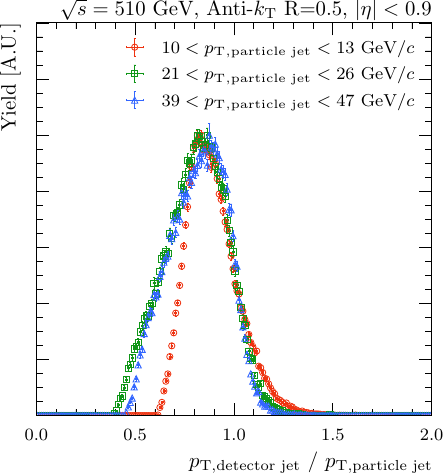}
\caption{\label{fig:ptratio} The distribution of reconstructed jets as a function of $\frac{p_{\text{T},\text{detector jet}}}{p_{\text{T},\text{particle jet}}}$ for various particle-jet $\pt$ bins from the embedding sample where the corresponding detector jets were required to satisfy the exclusive JP0 trigger for the lowest $\pt$ bin, exclusive JP1 for the intermediate $\pt$ bin, and exclusive JP2 for the highest $\pt$ bin. The exclusive-trigger selections were based on the choice made in this analysis to avoid double counting among three jet-patch-triggered events with increasing thresholds.
The left plot is for $\sqrt{s}=200$~GeV and the right plot is for $510$~GeV.
}
}

\newcommand\figurefive{%
\includegraphics{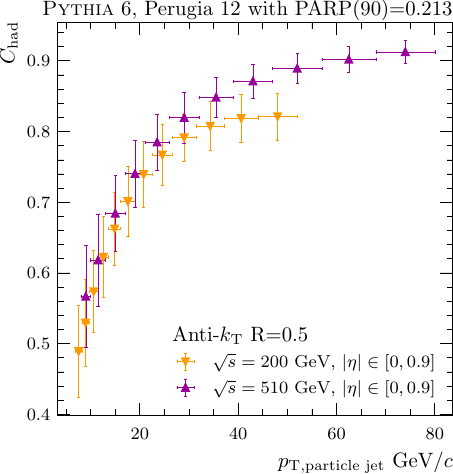}
\caption{\label{fig:chad} Hadronization correction factor at $\sqrt{s} = 200$ and $510$~GeV. Error bars represent systematic uncertainty obtained from Perugia 2012 tune variations which are related to alternative parameters for initial-state and final-state radiations, fragmentation process, and hadronization.
}
}

\newcommand\figureseven{%
\includegraphics{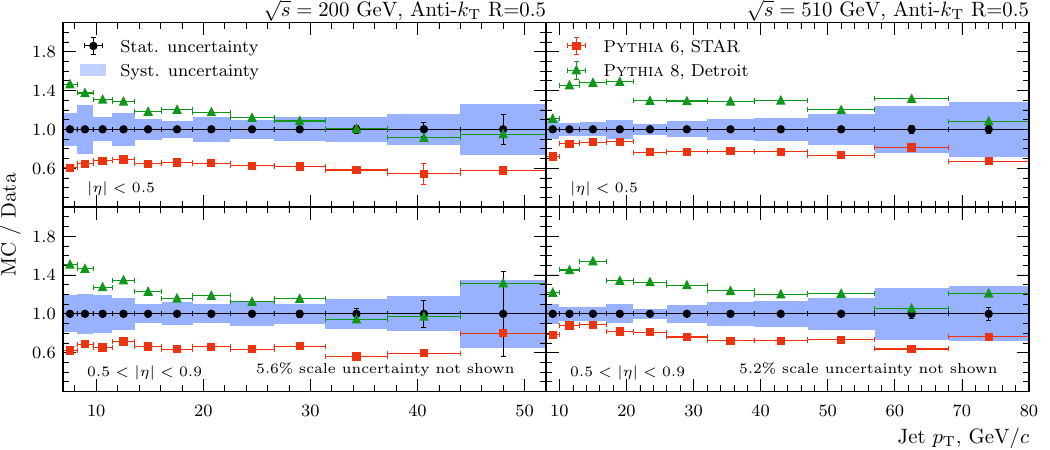}
\caption{\label{fig:cmppy} Ratio of jet cross sections predicted by \pythia{}~6 and \pythia{}~8 to the present measurements at $\sqrt{s} = 200$~GeV (left) and at $\sqrt{s} = 510$~GeV (right). Jets reconstructed from data and \pythia{} are both corrected for UE contributions using the off-axis cone method.
}
}

\newcommand\figurenine{%
\includegraphics{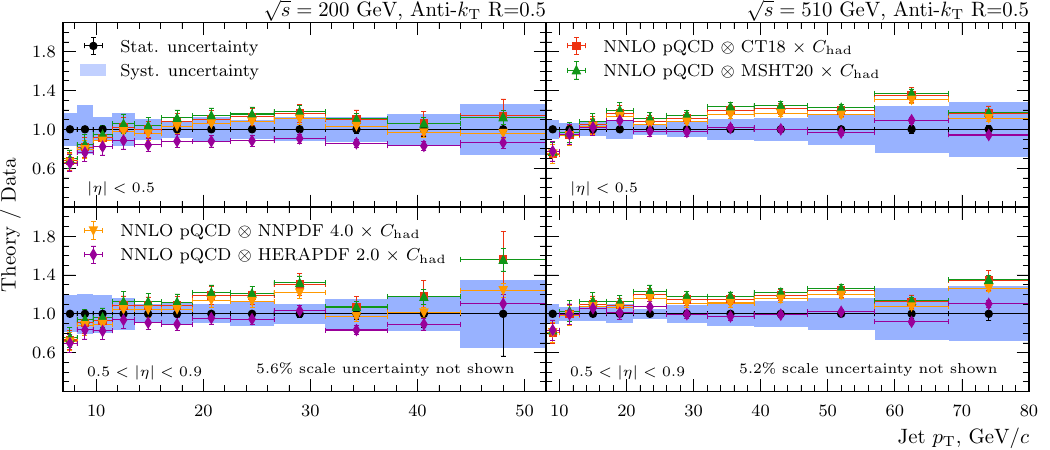}
\caption{\label{fig:cmpnnlo} Ratio of jet cross sections calculated from pQCD at NNLO with hadronization correction applied to cross sections from the present measurements at $\sqrt{s} = 200$~GeV (left) and at $\sqrt{s} = 510$~GeV (right). The NNLO PDFs include CT18 \cite{ct18}, MSHT20 \cite{msht20}, NNPDF4.0 \cite{nnpdf40} and HEARPDF2.0 \cite{hera20}.
}
}

\newcommand\figuretwelve{%
\includegraphics{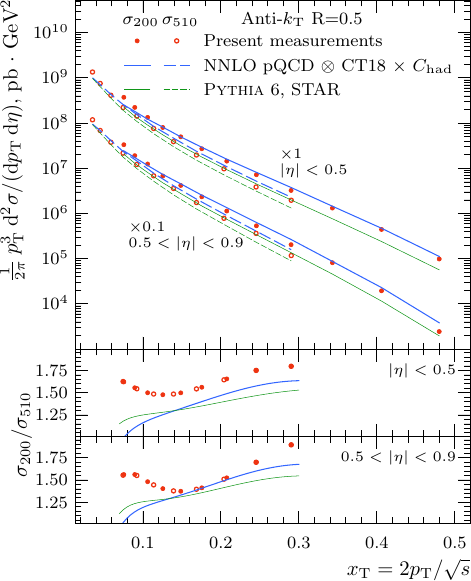}
\caption{\label{fig:cinv} Inclusive jet invariant jet cross section, $\frac{1}{2\pi}p^3_T\frac{d^2\sigma}{d\pt{}d\eta}$ vs. $\xt{}$ compared with NNLO pQCD calculations after hadronization corrections and \pythia{}~6 with STAR tune. Because the jet $\xt$ bins from the two measurements at $\sqrt{s} =$~200 and 510~GeV are not aligned, the ratios are based on the interpolated values from the inclusive jet cross sections at a given $x_T$ from one of the measurements. The ratios are for illustrative purposes; the uncorrelated and correlated uncertainties from the two measurements are not shown.
}
}

\newcommand\tableone{
\resizebox{0.48\textwidth}{!}{
    \begin{tabular}{|c|c|c|c|c|}
\hline
\multicolumn{5}{|c|}{$|\eta| < 0.5$} \\
\hline
Bin (GeV/$c$) & Total(\%) & Energy scale(\%) & Tracking efficiency(\%) & Unfold(\%) \\
\hline
$6.9$ -- $8.2$  & $17.1$        & $15.9$        & $2.1$ & $3.4$ \\
$8.2$ -- $9.7$  & $25.4$        & $24.6$        & $1.3$ & $5.3$ \\
$9.7$ -- $11.5$ & $12.5$        & $10.3$        & $2.4$ & $4.8$ \\
$11.5$ -- $13.6$        & $17.3$        & $16.6$        & $1.5$ & $3.9$ \\
$13.6$ -- $16.1$        & $10.7$        & $9.8$ & $2.2$ & $2.7$ \\
$16.1$ -- $19.0$        & $8.8$ & $7.9$ & $2.4$ & $2.1$ \\
$19.0$ -- $22.5$        & $13.0$        & $12.5$        & $2.5$ & $1.7$ \\
$22.5$ -- $26.6$        & $9.9$ & $9.2$ & $2.3$ & $1.4$ \\
$26.6$ -- $31.4$        & $12.3$        & $12.0$        & $3.4$ & $1.3$ \\
$31.4$ -- $37.2$        & $12.9$        & $12.6$        & $0.9$ & $1.4$ \\
$37.2$ -- $44.0$        & $16.2$        & $14.4$        & $5.1$ & $1.6$ \\
$44.0$ -- $52.0$        & $26.4$        & $25.3$        & $3.7$ & $2.1$ \\
\hline
\multicolumn{5}{|c|}{$0.5 < |\eta| < 0.9$} \\
\hline
Bin (GeV/$c$) & Total(\%) & Energy scale(\%) & Tracking efficiency(\%) & Unfold(\%) \\
\hline
$6.9$ -- $8.2$  & $19.1$        & $17.4$        & $1.7$ & $4.0$ \\
$8.2$ -- $9.7$  & $20.3$        & $19.1$        & $1.7$ & $6.7$ \\
$9.7$ -- $11.5$ & $19.4$        & $17.8$        & $2.0$ & $6.1$ \\
$11.5$ -- $13.6$        & $16.6$        & $14.3$        & $1.7$ & $5.4$ \\
$13.6$ -- $16.1$        & $9.9$ & $8.6$ & $2.4$ & $3.7$ \\
$16.1$ -- $19.0$        & $12.0$        & $11.1$        & $2.5$ & $2.8$ \\
$19.0$ -- $22.5$        & $9.8$ & $8.9$ & $1.9$ & $2.5$ \\
$22.5$ -- $26.6$        & $12.8$        & $12.1$        & $2.9$ & $2.3$ \\
$26.6$ -- $31.4$        & $10.5$        & $9.9$ & $2.7$ & $2.3$ \\
$31.4$ -- $37.2$        & $15.2$        & $14.9$        & $3.8$ & $2.2$ \\
$37.2$ -- $44.0$        & $17.9$        & $17.4$        & $1.2$ & $3.2$ \\
$44.0$ -- $52.0$        & $35.1$        & $34.0$        & $3.4$ & $6.1$ \\
    \hline
    \end{tabular}
    }
    \caption{Percent systematic uncertainties for inclusive-jet cross section at $\sqrt{s}= 200$~GeV.}
    \label{tab:syst200}
}

\newcommand\tabletwo{
\resizebox{0.48\textwidth}{!}{
    \begin{tabular}{|c|c|c|c|c|}
                \hline
                \multicolumn{5}{|c|}{$|\eta|<$ 0.5} \\
    \hline
            Bin(GeV/$c$)&           Total(\%)&    Energy scale(\%)&       Tracking efficiency(\%)&           Unfold(\%) \\
            \hline
       8 -- 10&             9.8&             7.3&             1.4&             6.3 \\
      10 -- 13&             6.9&             4.5&             1.1&             5.1 \\
      13 -- 17&             7.4&             4.8&             3.0&             4.7 \\
      17 -- 21&             9.5&             7.0&             4.1&             4.9 \\
      21 -- 26&             5.8&             3.3&             1.7&             4.4 \\
      26 -- 32&             8.4&             5.9&             3.1&             5.1 \\
      32 -- 39&            10.8&             8.5&             4.4&             5.0 \\
      39 -- 47&            12.1&             9.8&             5.1&             4.9 \\
      47 -- 57&            15.6&            12.7&             6.6&             6.3 \\
      57 -- 68&            23.9&            19.7&             9.7&             9.5 \\
      68 -- 80&            28.4&            22.7&            11.0&            13.0 \\
      \hline
                \multicolumn{5}{|c|}{0.5 $<|\eta|<$ 0.9} \\
    \hline
                  Bin(GeV/$c$)&           Total(\%)&    Energy scale(\%)&       Tracking efficiency(\%)&           Unfold(\%) \\
    \hline
      8 -- 10&             10.5&             8.5&             0.6&             6.2 \\
      10 -- 13&             7.0&             4.9&             0.8&             5.0 \\
      13 -- 17&             7.5&             4.9&             3.3&             4.5 \\
      17 -- 21&             9.9&             7.3&             4.4&             5.0 \\
      21 -- 26&             5.4&             2.7&             1.5&             4.5 \\
      26 -- 32&             9.1&             6.7&             3.6&             5.1 \\
      32 -- 39&            12.0&             9.5&             4.8&             5.6 \\
      39 -- 47&            13.5&            11.2&             5.5&             5.3 \\
      47 -- 57&            16.4&            13.7&             6.6&             6.2 \\
      57 -- 68&            27.1&            21.6&             9.9&            13.0 \\
      68 -- 80&            28.4&            24.2&            10.9&            10.3 \\
\hline
    \end{tabular}
    }
    \caption{Percent systematic uncertainties for inclusive-jet cross section at $\sqrt{s}= 510$~GeV.}
    \label{tab:syst510}
}

\newcommand\tablethree{
    \begin{tabular}{|c|c|}
\hline
\multicolumn{2}{|c|}{$|\eta| < 0.5$} \\
\hline
$(\pt{} - \Delta\pt^{\text{UE}})$ (GeV/$c$) &          $\frac{\mathrm{d}^2\sigma}{\mathrm{d}\pt\mathrm{d}\eta} (\text{pb}/\text{GeV})$ $\pm$ stat.(\%) $\pm$ syst. (\%) \\
\hline
$6.9$ -- $8.2$	& $5.42 \times 10^{6} \pm 0.4\% \pm 17.1\%$ \\
$8.2$ -- $9.7$	& $1.95 \times 10^{6} \pm 0.5\% \pm 25.4\%$ \\
$9.7$ -- $11.5$	& $7.03 \times 10^{5} \pm 0.5\% \pm 12.5\%$ \\
$11.5$ -- $13.6$	& $2.53 \times 10^{5} \pm 0.5\% \pm 17.3\%$ \\
$13.6$ -- $16.1$	& $9.51 \times 10^{4} \pm 0.6\% \pm 10.7\%$ \\
$16.1$ -- $19.0$	& $3.14 \times 10^{4} \pm 0.7\% \pm 8.8\%$ \\
$19.0$ -- $22.5$	& $1.01 \times 10^{4} \pm 1.0\% \pm 13.0\%$ \\
$22.5$ -- $26.6$	& $3.06 \times 10^{3} \pm 1.4\% \pm 9.9\%$ \\
$26.6$ -- $31.4$	& $8.31 \times 10^{2} \pm 2.3\% \pm 12.3\%$ \\
$31.4$ -- $37.2$	& $2.05 \times 10^{2} \pm 3.8\% \pm 12.9\%$ \\
$37.2$ -- $44.0$	& $4.15 \times 10^{1} \pm 7.3\% \pm 16.2\%$ \\
$44.0$ -- $52.0$	& $5.59 \times 10^{0} \pm 15.4\% \pm 26.4\%$ \\
\hline
\multicolumn{2}{|c|}{$0.5 < |\eta| < 0.9$} \\
\hline
$(\pt{} - \Delta\pt^{\text{UE}})$ (GeV/$c$) &          $\frac{\mathrm{d}^2\sigma}{\mathrm{d}\pt\mathrm{d}\eta} (\text{pb}/\text{GeV})$ $\pm$ stat.(\%) $\pm$ syst. (\%) \\
\hline
$6.9$ -- $8.2$	& $4.84 \times 10^{6} \pm 0.4\% \pm 19.1\%$ \\
$8.2$ -- $9.7$	& $1.67 \times 10^{6} \pm 0.6\% \pm 20.3\%$ \\
$9.7$ -- $11.5$	& $6.54 \times 10^{5} \pm 0.6\% \pm 19.4\%$ \\
$11.5$ -- $13.6$	& $2.14 \times 10^{5} \pm 0.7\% \pm 16.6\%$ \\
$13.6$ -- $16.1$	& $7.99 \times 10^{4} \pm 0.7\% \pm 9.9\%$ \\
$16.1$ -- $19.0$	& $2.71 \times 10^{4} \pm 1.0\% \pm 12.0\%$ \\
$19.0$ -- $22.5$	& $7.96 \times 10^{3} \pm 1.3\% \pm 9.8\%$ \\
$22.5$ -- $26.6$	& $2.26 \times 10^{3} \pm 2.0\% \pm 12.8\%$ \\
$26.6$ -- $31.4$	& $5.25 \times 10^{2} \pm 3.5\% \pm 10.5\%$ \\
$31.4$ -- $37.2$	& $1.27 \times 10^{2} \pm 5.7\% \pm 15.2\%$ \\
$37.2$ -- $44.0$	& $1.82 \times 10^{1} \pm 13.8\% \pm 17.9\%$ \\
$44.0$ -- $52.0$	& $1.39 \times 10^{0} \pm 43.5\% \pm 35.1\%$ \\
\hline
    \end{tabular}
    \caption{Double-differential inclusive-jet cross section at $\sqrt{s} = 200$~GeV.}
    \label{tab:sigma200}
}

\newcommand\tablefour{
    \begin{tabular}{|c|c|}
    \hline
    \multicolumn{2}{|c|}{$|\eta|<$ 0.5} \\
    \hline
               $(\pt{} - \Delta\pt^{\text{UE}})$ (GeV/$c$) &          $\frac{\mathrm{d}^2\sigma}{\mathrm{d}\pt\mathrm{d}\eta} (\text{pb}/\text{GeV})$ $\pm$ stat.(\%) $\pm$ syst. (\%) \\
\hline               8 -- 10&                                       1.16$\times 10^{7}$ $\pm$ 0.2\% $\pm$ 9.8 \%\\
              10 -- 13&                                       3.11$\times 10^{6}$ $\pm$ 0.1\% $\pm$ 6.9 \%\\
              13 -- 17&                                       7.65$\times 10^{5}$ $\pm$ 0.2\% $\pm$ 7.4 \%\\
              17 -- 21&                                       2.02$\times 10^{5}$ $\pm$ 0.4\% $\pm$ 9.5 \%\\
              21 -- 26&                                       6.89$\times 10^{4}$ $\pm$ 0.3\% $\pm$ 5.8 \%\\
              26 -- 32&                                       1.95$\times 10^{4}$ $\pm$ 0.4\% $\pm$ 8.4 \%\\
              32 -- 39&                                      5.45$\times 10^{3}$ $\pm$ 0.7\% $\pm$ 10.8 \%\\
              39 -- 47&                                      1.55$\times 10^{3}$ $\pm$ 1.1\% $\pm$ 12.1 \%\\
              47 -- 57&                                           434 $\pm$ 1.5\% $\pm$ 15.6 \%\\
              57 -- 68&                                          99.2 $\pm$ 4.0\% $\pm$ 23.9 \%\\
              68 -- 80&                                          30.4 $\pm$ 4.1\% $\pm$ 28.4 \%\\
\hline
    \multicolumn{2}{|c|}{0.5$<|\eta|<$ 0.9} \\
    \hline
            $(\pt{} - \Delta\pt^{\text{UE}})$ (GeV/$c$)&          $\frac{\mathrm{d}^2\sigma}{\mathrm{d}\pt\mathrm{d}\eta} (\text{pb}/\text{GeV})$ $\pm$ stat.(\%) $\pm$ syst. (\%)\\
\hline               8 -- 10&                                      1.02$\times 10^{7}$ $\pm$ 0.3\% $\pm$ 10.5 \%\\
              10 -- 13&                                       2.83$\times 10^{6}$ $\pm$ 0.2\% $\pm$ 7.0 \%\\
              13 -- 17&                                       6.99$\times 10^{5}$ $\pm$ 0.3\% $\pm$ 7.5 \%\\
              17 -- 21&                                       1.98$\times 10^{5}$ $\pm$ 0.5\% $\pm$ 9.9 \%\\
              21 -- 26&                                       5.87$\times 10^{4}$ $\pm$ 0.4\% $\pm$ 5.4 \%\\
              26 -- 32&                                       1.75$\times 10^{4}$ $\pm$ 0.6\% $\pm$ 9.1 \%\\
              32 -- 39&                                      5.01$\times 10^{3}$ $\pm$ 0.9\% $\pm$ 12.0 \%\\
              39 -- 47&                                      1.38$\times 10^{3}$ $\pm$ 1.4\% $\pm$ 13.5 \%\\
              47 -- 57&                                           349 $\pm$ 2.1\% $\pm$ 16.4 \%\\
              57 -- 68&                                          93.4 $\pm$ 4.5\% $\pm$ 27.1 \%\\
              68 -- 80&                                          18.1 $\pm$ 6.9\% $\pm$ 28.4 \%\\
\hline
    \end{tabular}
    \caption{Double-differential inclusive-jet cross section at $\sqrt{s} = 510$~GeV.}
    \label{tab:sigma510}
}

\begin{document}

\preprint{APS/123-QED}

\title{Inclusive jet cross section in $pp$ collisions at $\sqrt{s} = 200$ and $510$~GeV}

\affiliation{Academia Sinica, Nankang, 115}
\affiliation{Panchayat College (PC), Bargarh, affiliated with Sambalpur University (SU), Odisha 768028, India}
\affiliation{Abilene Christian University, Abilene, Texas   79699}
\affiliation{AGH University of Krakow, FPACS, Cracow 30-059, Poland}
\affiliation{Argonne National Laboratory, Argonne, Illinois 60439}
\affiliation{American University in Cairo, New Cairo 11835, Egypt}
\affiliation{Ball State University, Muncie, Indiana, 47306}
\affiliation{Brookhaven National Laboratory, Upton, New York 11973}
\affiliation{University of Calabria \& INFN-Cosenza, Rende 87036, Italy}
\affiliation{University of California, Berkeley, California 94720}
\affiliation{University of California, Davis, California 95616}
\affiliation{University of California, Los Angeles, California 90095}
\affiliation{University of California, Riverside, California 92521}
\affiliation{Central China Normal University, Wuhan, Hubei 430079 }
\affiliation{University of Illinois at Chicago, Chicago, Illinois 60607}
\affiliation{Chongqing University, Chongqing, 401331}
\affiliation{Creighton University, Omaha, Nebraska 68178}
\affiliation{Czech Technical University in Prague, FNSPE, Prague 115 19, Czech Republic}
\affiliation{Technische Universit\"at Darmstadt, Darmstadt 64289, Germany}
\affiliation{National Institute of Technology Durgapur, Durgapur - 713209, India}
\affiliation{ELTE E\"otv\"os Lor\'and University, Budapest, Hungary H-1117}
\affiliation{Frankfurt Institute for Advanced Studies FIAS, Frankfurt 60438, Germany}
\affiliation{Fudan University, Shanghai, 200433 }
\affiliation{Guangxi Normal University, Guilin, 541004}
\affiliation{University of Heidelberg, Heidelberg 69120, Germany }
\affiliation{University of Houston, Houston, Texas 77204}
\affiliation{Huzhou University, Huzhou, Zhejiang  313000}
\affiliation{Indian Institute of Science Education and Research (IISER), Berhampur 760010 , India}
\affiliation{Indian Institute of Science Education and Research (IISER) Tirupati, Tirupati 517507, India}
\affiliation{Indian Institute Technology, Patna, Bihar 801106, India}
\affiliation{Indiana University, Bloomington, Indiana 47408}
\affiliation{Institute of Modern Physics, Chinese Academy of Sciences, Lanzhou, Gansu 730000 }
\affiliation{University of Jammu, Jammu 180001, India}
\affiliation{Kent State University, Kent, Ohio 44242}
\affiliation{University of Kentucky, Lexington, Kentucky 40506-0055}
\affiliation{Lanzhou University, Lanzhou, 730000}
\affiliation{Lawrence Berkeley National Laboratory, Berkeley, California 94720}
\affiliation{Lehigh University, Bethlehem, Pennsylvania 18015}
\affiliation{Lovely Professional University, Jalandhar - Delhi G.T. Road, Pagwara, Panjab, 144411, India}
\affiliation{Max-Planck-Institut f\"ur Physik, Munich 80805, Germany}
\affiliation{Michigan State University, East Lansing, Michigan 48824}
\affiliation{National Institute of Science Education and Research, HBNI, Jatni 752050, India}
\affiliation{National Cheng Kung University, Tainan 70101 }
\affiliation{Nuclear Physics Institute of the CAS, Rez 250 68, Czech Republic}
\affiliation{The Ohio State University, Columbus, Ohio 43210}
\affiliation{Panjab University, Chandigarh 160014, India}
\affiliation{Purdue University, West Lafayette, Indiana 47907}
\affiliation{Rice University, Houston, Texas 77251}
\affiliation{Rutgers University, Piscataway, New Jersey 08854}
\affiliation{University of Science and Technology of China, Hefei, Anhui 230026}
\affiliation{South China Normal University, Guangzhou, Guangdong 510631}
\affiliation{Sejong University, Seoul, 05006, Korea, Republic Of}
\affiliation{Shandong University, Qingdao, Shandong 266237}
\affiliation{Shanghai Institute of Applied Physics, Chinese Academy of Sciences, Shanghai 201800}
\affiliation{Southern Connecticut State University, New Haven, Connecticut 06515}
\affiliation{State University of New York, Stony Brook, New York 11794}
\affiliation{Instituto de Alta Investigaci\'on, Universidad de Tarapac\'a, Arica 1000000, Chile}
\affiliation{Temple University, Philadelphia, Pennsylvania 19122}
\affiliation{Texas A\&M University, College Station, Texas 77843}
\affiliation{Texas Southern University, Houston, Texas, 77004}
\affiliation{University of Texas, Austin, Texas 78712}
\affiliation{Tsinghua University, Beijing 100084}
\affiliation{University of Tsukuba, Tsukuba, Ibaraki 305-8571, Japan}
\affiliation{University of Chinese Academy of Sciences, Beijing, 101408}
\affiliation{United States Naval Academy, Annapolis, Maryland 21402}
\affiliation{Valparaiso University, Valparaiso, Indiana 46383}
\affiliation{Variable Energy Cyclotron Centre, Kolkata 700064, India}
\affiliation{Warsaw University of Technology, Warsaw 00-661, Poland}
\affiliation{Wayne State University, Detroit, Michigan 48201}
\affiliation{Wuhan University of Science and Technology, Wuhan, Hubei 430065}
\affiliation{Yale University, New Haven, Connecticut 06520}

\author{B.~E.~Aboona}\affiliation{Texas A\&M University, College Station, Texas 77843}
\author{J.~Adam}\affiliation{Czech Technical University in Prague, FNSPE, Prague 115 19, Czech Republic}
\author{L.~Adamczyk}\affiliation{AGH University of Krakow, FPACS, Cracow 30-059, Poland}
\author{I.~Aggarwal}\affiliation{Panjab University, Chandigarh 160014, India}
\author{M.~M.~Aggarwal}\affiliation{Panjab University, Chandigarh 160014, India}
\author{Z.~Ahammed}\affiliation{Variable Energy Cyclotron Centre, Kolkata 700064, India}
\author{A.~K.~Alshammri}\affiliation{Kent State University, Kent, Ohio 44242}
\author{E.~C.~Aschenauer}\affiliation{Brookhaven National Laboratory, Upton, New York 11973}
\author{S.~Aslam}\affiliation{Fudan University, Shanghai, 200433 }
\author{J.~Atchison}\affiliation{Abilene Christian University, Abilene, Texas   79699}
% ...

\collaboration{STAR Collaboration}\noaffiliation

\noaffiliation

\date{\today}

\begin{abstract}
\ourabstract
\end{abstract}

\maketitle


\section{\label{sec:intro}Introduction}
The structure of the proton has been studied extensively over the past several decades. Deep inelastic scattering (DIS) data have revealed that the ordinary proton is a rich and complex bound state, nominally composed of two up quarks and one down quark held together by gluons, the carriers of the strong force, and the source of the sea of multiflavored quark-antiquark pairs that fluctuate on very short timescales~\cite{dis2012,hera20}. Based on the theory of quantum chromodynamics (QCD), partons are distributed according to the parton distribution function (PDF), $f_i(x, Q^2)$, parametrized by the fraction, $x$, of the hadron's longitudinal momentum carried by the parton being probed at the scale $Q^2$. The factorization framework allows us to connect perturbative partonic scattering amplitudes and nonperturbative PDFs to the observed scattering cross sections from particle colliders~\cite{Collins1989}.

The hard $2\rightarrow 2$ scattering with two incoming partons, each coming from one of the colliding protons, is an effective channel to probe the proton structure in proton-proton ($pp$) collisions~\cite{pdfReview2020}. The scattered partons from the hard interaction fragment then hadronize into collimated clusters of particles, known as jets~\cite{jets2008}. At the center-of-mass energies provided by the Relativistic Heavy Ion Collider (RHIC), $\sqrt{s} = 200$ and $510$~GeV, the produced jets are sensitive to gluons inside the protons~\cite{Mukherjee:nlo2012} and therefore provide direct access to the gluon distribution in a manner complementary to DIS experiments. Compared to previous inclusive jet cross section measurements at the TeV scale from the Tevatron~\cite{cdfprl1993,d0plb2002,tevatron2002,cdfprd2008,d0prl2008} and the Large Hadron Collider (LHC) \cite{ATLAS2013,ATLAS:jet8TeV2017,ATLAS:jet13TeV2017,ALICE:2.76TeVjet2013,CMS:jet2014,CMS:jetRatio2016,CMSPRC2017,CMS:13TeV2022}, measurements at RHIC energies are particularly sensitive to gluons with $x \gtrsim 0.1$. The momentum fraction, $x$, is proportional to a dimensionless variable, jet $\xt{} = 2\pt/\sqrt{s}$, for mid-rapidity jets, $\eta \sim 0$, where $\pt{}$ is the jet transverse momentum. The STAR Collaboration has reported the inclusive jet cross section in the jet pseudorapidity range $0.2 < \eta < 0.8$ at $\sqrt{s} =$~200 GeV from $pp$ collisions~\cite{STAR:2006opb}. The PHENIX Collaboration published their inclusive jet cross section measurement with $|\eta|< 0.15$ at $\sqrt{s} = 200$~GeV~\cite{phenix2024}.

Nearly five decades ago, Feynman, Field, and Fox proposed the following differential invariant cross section for meson production in hadron-hadron collisions $f(\Vec{p}) = \pt^4 E\frac{d^3\sigma}{d^3p}$ \cite{invCrss1978}. Integrating over the solid angle produces the expression $f(\xt{}) = \frac{1}{2\pi}\pt^3\frac{d^2 \sigma}{d\pt{}d\eta}$.  The invariant cross sections are expressed as a function of jet $\xt{}$, so that they can be compared at different values of $\sqrt{s}$~\cite{cdfprl1993,tevatron2002,ATLAS2013}. The ratios of invariant cross sections provide tests of the factorization and normalization scale dependence of the perturbative QCD theory. 

The underlying event (UE) associated with the parton hard scattering is composed of beam remnants and multiple parton interactions. The magnitude of the UE contributions at RHIC is different from that seen at TeV-scale hadron colliders~\cite{STAR:ue2019}.
For the Tevatron and LHC measurements, the UE contributions to the inclusive jet cross section are often included, together with the hadronization correction, as a part of a nonperturbative correction estimated using Monte Carlo (MC) event generator models.
In our analysis, we employ a data-driven UE correction method and estimate the hadronization correction with the MC-based approach.

\pythia{}~\cite{pythia6,pythia8:2022}, a modern event generator, uses a fixed-order matrix element from perturbative QCD calculations and parton-shower and hadronization models to simulate the hard scattering process in $pp$ collisions. It also incorporates sophisticated underlying event models. \pythia{} tunes take the measured inclusive jet cross sections at a wide range of $\sqrt{s}$ from experiments as the inputs to tune the parameters of their models. Several recent \pythia{} tunes used TeV-scale data as their reference. The behaviors at $\sqrt{s} =$~200~and~510~GeV are often extrapolated. A custom UE tune that uses the RHIC data as the baseline, the Detroit tune, is published~\cite{prd:detroit2021}. The inclusive jet cross sections at these energies will provide further constraints to those UE-related parameters and other $\sqrt{s}$-dependent parameters.




In this paper, we report the inclusive jet differential cross section as a function of jet transverse momentum, $\pt$, and pseudorapidity, $\eta$, at two center-of-mass energies, $\sqrt{s} = 200$ and $510$~GeV, measured by the STAR detector. The invariant jet cross sections, and their ratios, are presented at both energies as well. This is the first inclusive jet cross section measurement at $\sqrt{s} = 510$~GeV from STAR. A recently developed off-axis cone method to correct for the underlying event contribution to the measured jet $\pt$ is applied~\cite{star:jetALL2019}.

The paper is organized as follows: Section~\ref{sec:exp} will discuss the experimental apparatus, the sampled luminosity determination, jet reconstruction, and the technique used to correct for the underlying event. Section~\ref{sec:simu} focuses on the unfolding process used to correct triggered inclusive jet samples back to unbiased inclusive jet spectra by relying on simulations. Section~\ref{sec:syst} describes the contributions to the systematic uncertainty on the inclusive jet cross section measurement. Section~\ref{sec:hadcorr} discusses our suggested approach for correcting nonperturbative hadronization effects. Section~\ref{sec:results} presents data comparisons with model predictions, and conclusions are given in Sec.~\ref{sec:conclusion}.

\section{\label{sec:exp}Experiment and Data}

\subsection{\label{sec:star}The STAR detector at RHIC}
The results presented in this paper were obtained from data taken with the STAR detector at RHIC~\cite{rhic:design,rhic:proj,rhic:star} in 2012. RHIC is capable of colliding protons with center-of-mass energies between $62$ and $510$~GeV and is the only machine in the world capable of colliding polarized protons. RHIC consists of two rings. The rings circulate proton bunches in opposite directions. The bunch crossing frequency is approximately $9.38$~MHz. 

STAR is a suite of detectors, which includes a tracking system and electromagnetic calorimetry, located at the six-o'clock position of the RHIC ring. Charged particle tracking is provided by the Time Projection Chamber (TPC), which provides full azimuthal coverage and reasonable acceptance for tracks with $|\eta|$ of up to $\sim 1.3$~\cite{star:tpc}. The electromagnetic calorimeter system consists of a barrel calorimeter (BEMC) covering $-1< \eta < 1$ and an end-cap calorimeter  (EEMC) covering $1 < \eta < 2$ \cite{star:bemc,star:eemc}. Zero-degree calorimeters (ZDC) are located $18$~m away from the collision interaction point (IP) at $|\eta| > 6.6$. They detect photons and neutrons resulting from the collisions and serve as a beam luminosity monitor \cite{star:zdc}.

\subsection{\label{sec:lumi}Luminosity measurement}
A Van der Meer scan is a widely used method to determine the collision luminosity, which depends on the spatial distribution of protons in the colliding beams~\cite{lum:vs}. The luminosities based on this method have been used at STAR to measure, for example, the total and elastic $pp$ cross sections at both $\sqrt{s} =$~200~and 510~GeV~\cite{STAR:elastic200in2020,STAR:elastic510in2023}, as well as $W^{\pm}$ and $Z^0$ cross sections from $pp$ collisions at $\sqrt{s} =$~510~GeV~\cite{STAR:wz2020,STAR:z2024}. The method measures the changes in collision rate as a function of beam displacement along the horizontal or vertical direction in the plane perpendicular to the beam axis. These dedicated scans were taken for each specific $\sqrt{s}$. An effective cross section was calculated using a counting detector and applied to scale the measured collision rates during data-taking to obtain the sampled luminosity.

Two ZDCs, one on each side of STAR, were used as monitoring detectors. The ZDC coincidence rate, which required coincident hits above threshold values in both ZDCs, was chosen as the luminosity monitoring rate. It was corrected for accidental and multiple collisions~\cite{cdflumi2000}. The correction to the luminosity measurement due to varying transverse beam size along the beam moving direction, also known as the hourglass effect~\cite{lum:hg}, was 1$\%$-2$\%$. It was suppressed due to the fact that the beta function at the IP, $\beta^{*} \approx 1$~m, was comparable to the bunch length, $\sim$ $60$~cm.

The dominant systematic uncertainty in the luminosity, of roughly $3\%$, was due to the beam position measurements. Other contributing sources included fluctuations in the number of protons in a bunch and the small but nonzero crossing angles of the colliding  bunches. The total systematic uncertainties were found to be $5.6\%$ and $5.2\%$ for $\sqrt{s}=200$ and $510$~GeV, respectively.

The dead time for the STAR trigger system, normally between $5\%$ and $10\%$, was factored into the sampled luminosity calculation. After data quality assurance checks, the integrated luminosities of the analyzed data were 17 and 42 $\text{pb}^{-1}$ at $\sqrt{s}=$~200~and~510~GeV, respectively.

\subsection{\label{sec:trig}Triggered events}
Signals from both the BEMC and EEMC are fed into the STAR trigger system to select events of interest for this analysis. In the trigger logic, the BEMC and EEMC are divided into six adjacent sections in $\varphi$, and five overlapping sections in $\eta$: $\eta \in [-1, 0]$, $[-0.6, 0.4]$, $[0, 1]$, $[0.4, 1.4]$, and $[1, 2]$. Each of these fixed $1 \times \pi/3$ regions in $\eta$-$\varphi$ space is called a jet patch (JP). A JP event is triggered by requiring the sum of the analog-to-digital converters (ADCs) from a JP to be above a certain threshold. During the $200$~GeV run, three thresholds corresponding to transverse energies of 3.5, 5.4, and 7.3 GeV  were used for the JP0, JP1 and JP2 triggers. In the $510$~GeV run, these threshold energies were raised to 5.4, 7.3, and 14.4~GeV, respectively. The rates of accepted JP0- and JP1-triggered events are controlled by applying prescale factors during data collection in order to remain within the bandwidth of the STAR data-acquisition system and to balance the number of events across the jet-patch-triggered samples.

%
\subsection{\label{sec:jetrec}Jet reconstruction}
Jets were reconstructed from TPC tracks and BEMC and EEMC tower energy deposits, which were passed to the  anti-$\kt$ jet-finding algorithm implemented in FastJet~\cite{antikt,fastjet}. For the $\sqrt{s} = 510$~GeV dataset, the jet-resolution parameter, $R$, was optimized at a value of $0.5$ to reduce susceptibility to the soft diffuse background and pileup events, which increase at larger center-of-mass energies~\cite{star:jetALL2019}. To facilitate comparisons, the $R$ parameter at $\sqrt{s} = 200$~GeV was kept consistent between the two datasets. The TPC only provided information on charged-particle momenta, so the pion mass was assumed for all tracks for the purpose of constructing the 4-momenta that were passed to the jet finder. Similarly, the BEMC and EEMC towers measured only deposited energy, and the resultant 4-momenta were reconstructed assuming massless particles with trajectories originating at the collision vertex. 

The selected TPC tracks were required to have at least 12 hit points out of a possible 45 to provide good momentum resolution. To remove split tracks, the number of hit points was also required to be greater than $51\%$ of the maximum possible number of hit points when the track geometry and active electronic channels were considered.  In addition, the tracks were required to have transverse momenta $\pt > 0.2$~GeV/$c$ and be associated with the selected collision vertex for the event within a $\pt$-dependent distance of closest approach (DCA). The DCA was required to be less than 2~cm for $\pt < 0.5$~GeV/$c$ and less than 1~cm for $\pt > 1.5$~GeV/$c$; the DCA requirement was linearly interpolated in the region between these two limits. The BEMC and EEMC tower hits were required to have an ADC signal well above the pedestal and the calibrated transverse energy $\Et \equiv E \sin\theta > 0.2$~GeV, where $\theta$ is the polar angle of a tower in the STAR lab frame. 

If a track could be projected onto a BEMC/EEMC tower, the track transverse energy $\pt \cdot c$ was subtracted from the matched tower $\Et$.  If the difference was negative, the contribution from that tower was discarded. This procedure, referred to as ``$\pt$ subtraction,'' suppressed double-counting by EMC towers from fully reconstructed TPC tracks. Hadrons could leave energy in a cluster of towers, in which case the $\pt$ subtraction resulted in a systematic undersubtraction. In the rare case (about 1\% of charged hadrons) where a photon struck the same tower as a charged hadron, an oversubtraction happened. Overall, the $\pt$ subtraction improved the resolution of the reconstructed jet energy by suppressing the contribution of the charged-hadron energy deposition in the electromagnetic calorimeters.

A pileup-proof vertex finder~\cite{starPPV2010} was used to reconstruct collision vertices. In this analysis, an event was discarded if no good vertex was found. The selected vertex was the one least susceptible to pileup events for events that passed. By requiring all TPC tracks to be associated with the best vertex and to satisfy the DCA cut, we removed most of the backgrounds, such as pileup, beam-gas, and charged cosmic rays. Furthermore, the jets analyzed were required to contain at least one track. This left the underlying event background as the major background we focused on.

\subsection{\label{sec:ue}Underlying event correction}
Although the UE contribution is expected to be uniformly distributed in $\eta$-$\varphi$ space, the STAR detector has a reasonably uniform acceptance in azimuth, but not in pseudorapidity. An $\eta$-dependent UE correction, called the ``off-axis cone'' method~\cite{Abelev2015,ALICE:cjet2018}, was introduced to fix the nonuniformity in $\eta$ of the STAR detector. This technique was first applied in the inclusive jet and dijet longitudinal double-spin asymmetry analysis at STAR~\cite{star:jetALL2019}. For a given jet, the procedure was to look at the particles measured in the two off-axis cones centered at $\pm 90^{\circ}$ away in $\varphi$ and at the same jet $\eta$. The averaged transverse momentum density was calculated for both cones, $\rho_{\text{cone }i}$, along with the jet area, $A_{\text{jet}}$, to define a shift, $\delta p_{T} = A_{\text{jet}} (\rho_{\text{cone }1} + \rho_{\text{cone }2}) / 2$, that was subtracted from the raw reconstructed jet $\pt$. The $A_{\text{jet}}$ was calculated by the FastJet package \cite{fastjet} using their default ghost particle method.

The underlying event correction was at the level of $1$~GeV/$c$ and mostly flat across the measured jet $\pt$ ranges. The energy density extracted from the off-axis cone analysis was anisotropic. Both the data and embedding samples demonstrated an average 5\% increase of the density on the side of the off-axis cone away from the jet compared to the side near the jet. Additionally, a roughly 15\% difference was found in the underlying event $\pt{}$-correction between data and simulation. This difference was much smaller than the systematic uncertainties due to the detector effects and the unfolding discussed below in Sec.~\ref{sec:syst}. Therefore the uncertainty due to underlying event correction on the jet cross section was negligible.


\subsection{\label{sec:counting_jets}Counting jets}
 An event that passed a higher JP threshold had automatically satisfied the condition for a lower threshold, and therefore, depending on the prescale values, as mentioned in Sec.~\ref{sec:trig}, an event could be identified as firing more than one JP trigger. To avoid statistical correlations among the overlapping triggered samples, an {\it exclusive} jet-triggered sample was formed by first assigning each event to only one trigger group based on the highest applicable threshold and then combining all the trigger groups. 

Jets reconstructed from the JP-triggered events were ``geometrically matched'' by requiring them to point to a jet patch with a sum of ADC signals above the corresponding trigger threshold. Jets were binned based on their UE-subtracted transverse momenta. The intervals of selected jet $\pt$ values were
$[6.9, 25]$, $[8, 52]$, and $[9.5, 52]$~GeV/$c$
($[8, 33.6]$, $[10, 39.3]$, and $[21, 80] $~GeV/$c$) 
for JP0, JP1 and JP2 triggered events for the $\sqrt{s} = 200$~GeV ($\sqrt{s} = 510$~GeV) sample. The two pseudorapidity ranges were $|\eta| < 0.5$ and $0.5 < |\eta| < 0.9$ for all triggers.

\section{\label{sec:simu}Simulation}
Computer simulation is an essential component of this analysis, used for both correcting the raw data to account for detector effects and evaluating systematic uncertainties. The Monte Carlo event generator also provides a means to study the hadronization correction needed to relate the experimentally determined particle-level cross section to a fixed-order pQCD calculation.

\subsection{\label{sec:embed}Embedding process}
The simulation sample was generated using a process called embedding, where detector signals recorded for zero-bias events were embedded into a full detector simulation of events produced by \pythia{}. Zero-bias events were collected without any trigger requirement on one out of a fixed number of bunch crossings during experimental data taking. Events were simulated using \pythia{} version 6.4.28~\cite{pythia6} and the Perugia 2012~\cite{Skands:2010ak} tune with the $\text{PARP}(90)$ value lowered from the default value of 0.24 to 0.213. This setting has been shown to match STAR's previously published identified hadron invariant yields from $pp$ collisions at $\sqrt{s} = 200$~GeV~\cite{starpipm2006,starpipm2012} well. Detector responses were simulated via a model of the STAR detector based on \textsc{GEANT3}~\cite{geant3}.


\subsection{\label{sec:unfolding}Unfolding}
The unfolding process included three elements: matching ratios, the unfolding matrix, and efficiencies. Matching ratios, sometimes referred to as background ratios, accounted for the fraction of detected jets not associated with particle jets. The detector jets are those reconstructed from TPC tracks and EMC towers. The particle jets are those clustered from the final-state generated particles in MC events without detector-response simulation. These ratios were estimated from simulation by requiring a matching criterion between detector jets and particle jets. The distance in $\eta$-$\varphi$ space between a detector jet and particle jet was defined as $\Delta R = \sqrt{\Delta \eta^2 + \Delta \varphi^2}$. A match was found when the minimum distance, $\Delta R_{\text{min}} < R$, where $R$ was the jet-resolution parameter used in the jet-finding process. The matching ratios were close to 1, except at the edges of the fiducial acceptance, where the values dipped to $\sim70\%$. 


The unfolding matrix, $A$, with rows representing detector jets and columns representing particle jets in two-dimensional kinematic bins, $p_{T}$ and $\eta$, were reconstructed from matched detector- and particle-jet pairs. The columns of $A$ were normalized by the number of matched pairs, so that $A_{i,j}$ represented the probability of a particle jet in $j$th bin being detected in the $i$th detector-jet bin. The particle-jet spectrum, $x$, was calculated directly by inverting $A$ and multiplying by the detector-jet spectrum $b$: $x = A^{-1}b$. $A$ can be visualized in terms of ratios of detector- and particle-jet $\pt{}$, $p_{T,\text{detector jet}}/p_{T,\text{particle jet}}$, as shown in Fig.~\ref{fig:ptratio}. Events for which this ratio is above $1$ are dominated by the detector resolution. Those for which it is below $1$ are affected by detector acceptance and efficiency, where in the smallest jet $\pt{}$ range they are also subject to threshold cutoffs.

\begin{figure*}[tb]
\figurefour
\end{figure*}%

The statistical uncertainty in the detector-jet spectrum was propagated through the unfolding process. Because the uncertainty in any unfolded particle-jet bin contained contributions from multiple detector-jet bins, small statistical uncertainties in the measured spectrum could be translated into large uncertainties on the unfolded spectrum. To overcome this problem, detector-jet bin sizes were chosen carefully to guarantee an acceptable unfolded uncertainty.

Efficiencies, $\varepsilon$, were used to correct for the fact that cuts were employed to select events and jets within a reasonable acceptance and to remove those events that might be poorly reconstructed or overly sensitive to pileup contributions or trigger biases. The particle-jet spectrum from simulations, without any detector-jet-matching requirement applied, was used as the reference spectrum to be compared with the matched particle-jet spectrum found with the unfolding matrix.

\section{\label{sec:syst}Systematic uncertainties}
The systematic uncertainties in the inclusive jet cross section are separated into two categories: those associated with detector effects, and those inherent to the unfolding process. Uncertainties associated with detector effects relate to how well the jet $\pt$ is measured in the data, with sources including the jet energy scale, and the tracking efficiency uncertainty. The unfolding uncertainty is driven by statistical limitations and the prior bias from the simulation. Each contribution is discussed below.



\subsection{\label{sec:detunc}Detector effect uncertainty}
The jet energy scale uncertainty had contributions from both the TPC track momentum calibration and the EMC energy scale. The global TPC track momentum calibration uncertainty was estimated to be $1\%$. The response of calorimeters to electromagnetic particles (such as electrons, positrons, and photons) and hadrons contributed to the energy scale uncertainty. For electromagnetic energy deposited in the calorimeter towers, the uncertainty of $3.8\%$ was determined by a gain calibration. The total hadronic activity in the calorimeters was estimated from the number and momenta of the TPC tracks. The TPC track contributions were weighted by the track reconstruction efficiency and boosted by a scale-up factor that accounts for the energy of the neutral hadrons not reconstructed by either the TPC or the calorimeters. The average TPC track efficiency is $90\%$ at $\sqrt{s} = 200$~GeV  and $75\%$ at $\sqrt{s} = 510$~GeV due to a large amount of pileup. The scale-up factor was taken as $1.16$~\cite{STAR:starEt2004} for both $\sqrt{s} = 200$~and~$510$~GeV. 

The uncertainty of hadron energy loss in the EM towers was studied from the simulation. The hadron shower model used in the simulation was the GHEISHA cascade model \cite{geisha1985} with a transport cut threshold at $1$~MeV, below which a showering track would deposit all of its remaining energy in the calorimeter tower.

The average fractional energy a hadron deposits into the calorimeters, $f_{h}$, was estimated by measuring charged-hadron tracks from the TPC, projecting them to a BEMC tower, and finding the total energy, $E_{\text{clust.}}$, in the 3$\times$3 cluster centered around the projected tower. Thus, $f_{h} = E_{\text{clust.}} / p$, where $p$ is the hadron's momentum. In order to select hadron tracks, and not electrons or positrons, the tracks were required to satisfy $\text{n}\sigma_{e} > 2$ or $\text{n}\sigma_{e} < -1$, where $\text{n}\sigma_{e}$ represented the deviation of the energy deposited by the track in the TPC from that expected for an identified electron. An isolation cut was also applied so that no other measured TPC track intersected the selected cluster. We found the average $\langle f_{h} \rangle$, weighted by the track momentum,
was around 32\%. The relative uncertainty, which was the difference of $\langle f_{h} \rangle$ between data and simulation, was 6\%. The $f_{h}$ distributions for charge-separated tracks with momentum between 2 and 3 GeV from data and simulations are shown in Fig.~\ref{fig:ecpvspt}, where the data are samples from the 510 GeV $pp$ collisions. The $\pt{}$ subtraction implemented in the jet reconstruction suppressed the contribution of hadron energy loss uncertainty to the measurement of the tower energies by 72\%; therefore, the remaining uncertainty for the EM tower energy due to unsubtracted hadronic energy was 1.6\%. 

\begin{figure}[tb]
\figurethree
\end{figure}%


The effect of the jet energy scale uncertainty on the detector-jet spectrum was estimated by shifting the jet $\pt$ up and down separately. The final impact on the particle-jet spectrum was determined by applying this shifted jet spectrum from data, and then unfolding the spectrum with the nominal unfolding matrix. The larger of the differences between the nominal unfolded spectrum and either the up-shifted or down-shifted spectrum was taken as the systematic uncertainty on the unfolded particle level jet cross section. At large jet $\pt{}$, this contribution reached 20\% or more and was the dominant systematic uncertainty.

The relative tracking efficiency uncertainty at $\sqrt{s} = 200$~GeV was about 1\%~\cite{startrkeff}. Due to larger track multiplicities and a greater number of soft tracks from pileup events at $\sqrt{s} = 510$ GeV, a $4\%$ relative tracking efficiency uncertainty was conservatively estimated~\cite{startrkeff}. To estimate the tracking efficiency uncertainty on the jet cross section, we resorted to the embedding sample. Before reconstructing the jets following the standard procedure, $1\%$ and $4\%$ of reconstructed TPC tracks were randomly rejected at $\sqrt{s} = $~200 and 510~GeV, respectively. Then we used the reconstructed jet spectra as our input to the unfolding process. The differences between the unfolded particle-jet spectra and the original unbiased particle-jet spectra were taken as the uncertainties due to the tracking efficiency. At $\sqrt{s} = 200$~GeV, the maximum uncertainty on the unfolded particle-jet spectrum was $5\%$. At $\sqrt{s} = 510$~GeV, the uncertainty was less than $5\%$ at low jet $\pt$, and more than $10\%$ at high jet $\pt$.

\subsection{\label{sec:unfold_unc}Unfolding uncertainty}
The systematic uncertainty from the unfolding process has two contributions: one related to simulation statistics and the other to prior bias. A conventional bootstrap method was used to estimate the systematic uncertainty due to the limited simulation sample~\cite{10.1214/aos/1176344552}. 


As discussed in Sec.~\ref{sec:embed}, events used to seed the \geant{} detector response simulation were generated with a STAR-specific tune for the \pythia{} generator. Differences in the trigger or detector response to jets arose from different subprocesses (e.g., by virtue of gluon jets being wider and containing more particles than the quark jets). This contribution was estimated by varying the initial input PDFs for the \pythia{} generator. This was achieved by reweighting the simulation based on the input PDF. One hundred replicas from the unpolarized NNPDF 3.0 set, in addition to the central PDF, were included in this analysis. An event weight for the $i$th PDF replica was calculated as $w_i=\frac{f_{1,i}f_{2,i}}{f_1f_2}$, where $f_{1(2),i}$ was the parton density for incoming parton 1(2) taken from the $i$th PDF replica and $f_{1(2)}$ was the parton density for incoming parton 1(2) from the nominal PDF used in the simulation. The reweighted detector-jet spectra were unfolded with the nominal unfolding matrix. The averaged difference between the unfolded particle-jet spectra and the reweighted particle-jet spectra was taken as the systematic uncertainty due to prior variation, and was found to be negligible.


Finally, the individual contributions were added in quadrature as the total systematic uncertainties, as summarized in Tables~\ref{tab:syst200}~and~\ref{tab:syst510}, for $\sqrt{s} =$~200 and 510~GeV, respectively. The dominant systematic uncertainty came from the jet energy scale. The subdominant contributions were from the tracking efficiency and the limited simulation sample size.

\begin{table}[hptb]
    \centering
    \tableone
\end{table}
\begin{table}[hptb]
    \centering
    \tabletwo
\end{table}

\section{\label{sec:hadcorr}Hadronization correction}
The cross section extracted from data was presented at the particle level, while the fixed-order pQCD calculation was only at the parton level. To bridge the two, a hadronization correction was applied to the pQCD calculation to allow a consistent comparison with our results. Fortunately, the present hadronization model provided by \pythia{} allowed us to quantify the size of this correction.

The hadronization correction was treated as a bin-by-bin multiplicative correction to be applied to the pQCD theoretical prediction and was found by taking the ratio of the particle-level jet cross section over the parton-level jet cross section in a common jet $\pt$ and $\eta$ bin, $C_{\text{had}} = \frac{\sigma_{\text{particle}}}{\sigma_{\text{parton}}}$. The bin-by-bin correction was sufficient, since \pythia{} reproduced the shape of the unfolded particle-jet distribution as a function of $\pt$ in the studied $\eta$ range. The parton-level jets were reconstructed from intermediate partons observed in \pythia{} right before the hadronization step. For parton and particle jets, the same off-axis cone correction was employed as for the data. Our hadronization correction had a larger magnitude compared to the traditional nonperturbative correction because the latter included a correction for the UEs that coincidentally was of a similar magnitude as hadronization, but of an opposite sign.


The uncertainty on $C_{\text{had}}$ was estimated from several variants of the Perugia 2012 tune with modified parameters for initial-state radiation (ISR) and final-state radiation (FSR), the fragmentation process, and the hadronization. The tunes regarding the ISR and FSR parameters were a pair of variations with $\alpha_s(\frac{1}{2}\pt)$ and $\alpha_s(2\pt)$, respectively, for both ISR and FSR. They captured the effect of the change in the parton-level jet cross sections on $C_{\text{had}}$. The variations with the fragmentation process were a pair of tunes with more longitudinal and more transverse fragmentation than the default tune, and the Innsbruck hadronization tune.  For the two pairs of tunes, the larger difference of each pair from the default was assigned as the uncertainty. The differences of the three types of variation to the default tune were summed in quadrature as the $C_{\text{had}}$ uncertainty. The hadronization correction factors, $C_{\text{had}}$, and their uncertainties are shown in Fig.~\ref{fig:chad}.

Note that for a given jet $\pt$, $C_\text{{had}}$ was slightly larger at higher $\sqrt{s}$. For example, at jet $\pt =$ 40 GeV, $C_{\text{had}}$ was close to 90\% at $\sqrt{s} = 510$, while at $\sqrt{s} =$ 200 GeV, $C_{\text{had}}$ was around 80\%. This discrepancy could be explained using an interpretation of the hadronization process as a shift of the jet $\pt$ from the parton level to the particle level. The first derivative of the jet cross section as a function of jet $\pt$ should be considered. Specifically, for a fixed $\pt$ shift at a given $\pt$ bin, the steeper slope of the jet $\pt$ spectra at $200$~GeV compared to $510$~GeV would translate to a larger deviation of $C_{\text{had}}$ from $1$. The assumption that the $\pt$ shifts at both collision energies were quite close (at around 1 GeV/$c$) was based on predictions by \pythia{}.

\begin{figure}[hptp]
\figurefive
\end{figure}%

\section{\label{sec:results}Results}

The double-differential inclusive-jet cross section, $\frac{\mathrm{d}^2\sigma}{\mathrm{d}\pt\mathrm{d}\eta}$, is presented as a function of jet $\pt$ within two $\eta$ ranges, $|\eta|<$ 0.5 and 0.5 $<|\eta|<$ 0.9 at $\sqrt{s} = 200$~and 510 GeV, as shown in Fig.~\ref{fig:pp_xsec} and Tables~\ref{tab:sigma200} and~\ref{tab:sigma510}, respectively. The results were obtained by combining data from three jet-patch triggers, JP0, JP1, and JP2. All of the unfolded results are unbiased particle-jet cross sections corrected with the matching ratios, unfolding matrix, and efficiencies calculated using the embedding sample.

\begin{figure}[hptb]
\figureone
\end{figure}%

The inclusive jet cross sections from \pythia{} event generators, which incorporate the Lund string hadronization model and provide the stable final-state particles, are compared to our results in Fig.~\ref{fig:cmppy}. The same jet-finding algorithm with the identical parameters and kinematic cuts as in the data analysis were applied to the final particles generated by \pythia{}. The reconstructed jet $p_T$ from the event generator was corrected by the off-axis cone UE method. Although the embedding was generated from the \pythia{}~6 Perugia~2012 tune with the modified $\text{PARP}(90)$ parameter, the results consistently underpredict the data by almost 30\%. In contrast, the newer ``Detroit'' tune of  \pythia{}~8~\cite{prd:detroit2021} overestimates the jet cross section by $20\%$-$40\%$, depending on the jet $\pt$. 

\begin{figure*}[hptb]
\figureseven
\end{figure*}%

Predictions at NNLO accuracy came from the NNLOJet theoretical framework~\cite{Gehrmann:2018szu,PhysRevLett.118.072002} with full color coherence in the form of FastNLO interpolation grids~\cite{applfast} that were produced with a recommended renormalization and factorization scale choice, $\mu = \hat{H}_{\text{T}}$~\cite{Currie2018}. In order to compare our results with these fixed-order pQCD calculations, the hadronization correction factor $C_{\text{had}}$, as discussed in Sec.~\ref{sec:hadcorr}, was applied to the NNLO calculations. Figure~\ref{fig:cmpnnlo} shows comparisons of these results with NNLO calculations using CT18~\cite{ct18}, HERAPDF2.0~\cite{hera20}, MSHT20~\cite{msht20}, and NNPDF4.0~\cite{nnpdf40} PDFs with the hadronization correction applied. For CT18, MSHT20, and NNPDF4.0 PDF predictions, our measured jet cross sections are larger than the predicted value at low jet $\pt{}$ and smaller by an almost constant 20\% at high jet $\pt$. It is interesting to note that our data agree well with the HERAPDF2.0 prediction except in the first few low jet-$\pt{}$ bins. The overall discrepancies shown in Fig.~\ref{fig:cmpnnlo} might not be explained solely from the respective PDF uncertainties, but also from missing higher-order contributions beyond the NNLO. Accurate estimations of uncertainties due to missing higher-order contributions or future developments on the next-to-next-to-next-to-leading-order calculations and beyond could accommodate the difference seen in Fig.~\ref{fig:cmpnnlo}.


While the PHENIX Collaboration also reported an inclusive jet cross section at $\sqrt{s} = 200$~GeV that was smaller than NNLO pQCD calculations~\cite{phenix2024}, several differences in the experimental results should be recognized. PHENIX presented an inclusive jet cross section in a limited $\eta$ range, $|\eta| <$ 0.15. Their results used a smaller jet-resolution parameter, $R=0.3$, for the anti-$\kt$ jet-finding algorithm rather than the $R=0.5$ chosen in this analysis. A larger $R$ value would capture more parton radiation, reduce hadronization corrections, be prone to UE contributions, and therefore result in an increase in the jet cross section~\cite{npC:2007wa}. Experiments at the LHC showed that the larger the value of $R$ is, the better the comparison between the data and the pQCD calculations after nonperturbative corrections~\cite{CMS:jet2014,CMSPRC2017}. Another difference is that PHENIX treated the UE contribution to the jet $\pt$ as a part of the nonperturbative correction and its uncertainty. 

\begin{figure*}[htb]
\figurenine
\end{figure*}%
\begin{table}
    \centering
    \tablethree
\end{table}

\begin{table}
    \centering
    \tablefour
\end{table}

Figure~\ref{fig:cinv} compares the invariant cross sections for inclusive jets at $\sqrt{s}=200$ and $510$ GeV as a function of $\xt{}$. Scale-breaking effects, i.e., deviations from unity are observed in the RHIC data, similar to effects reported from data collected at the Tevatron \cite{cdfprl1993,tevatron2002} and the LHC \cite{ATLAS2013}. The cross-section ratios at $\sqrt{s}=$~630 and 1800~GeV from the Tevatron span a similar kinematic range to those from the LHC, where the $\xt{}$ at $\sqrt{s}=$~2.76 and 7~TeV extend to lower values. We also show in Fig.~\ref{fig:cinv} that the NNLO pQCD with CT18 PDF calculations predict smaller ratios than what we have measured, especially at $\xt{} < 0.14$, but very similar trends at $\xt{} > 0.14$. The cross-section ratios vary with the renormalization and factorization scale dependence of $\alpha_S$ and PDFs used in pQCD calculations, instead of the choice of PDF. The ratios from \pythia{}~6 with the STAR tune lie close to the NNPDF pQCD calculation. It should be noted that Fig.~\ref{fig:cinv} offers a direct comparison of the inclusive jet cross sections at $\sqrt{s}=$~200 and 510 GeV; the correlated systematic uncertainties and the uncorrelated statistical uncertainties between the two measurements are not shown on the invariant jet cross section ratios, because the jet $\pt$ bins used in two measurements do not align with jet $\xt$ bins.

\begin{figure}[htb]
\figuretwelve
\end{figure}%

\section{\label{sec:conclusion}Conclusion}
The double-differential inclusive-jet cross sections, $\frac{\mathrm{d}^2\sigma}{\mathrm{d}\pt\mathrm{d}\eta}$, are presented as a function of jet transverse momentum, $\pt$, in two pseudorapidity ranges, $|\eta|< 0.5$ and $0.5 <|\eta| < 0.9$, at center-of-mass energies $\sqrt{s}=200$ and $510$ GeV. The off-axis cone method was applied to account for the underlying-event contributions to the jet $\pt{}$. This enabled us to separate corrections for the underlying event and hadronization effects. This approach is an improvement over previously published jet cross section analyses in high-energy $pp$ ($p\bar{p}$) collisions where the two corrections are treated as one.

The predictions from simulation event generators such as \pythia{}~6 and 8 were compared to our results. We found that \pythia{}~6 described the shape of the jet cross section well, while a scale difference was observed for both versions of \pythia{}. These results should provide crucial input for tuning the event generators at RHIC energies, especially parameters that depend on $\sqrt{s}$.

The measured particle-jet cross sections and the hadronization correction factor provided in this analysis permit a direct comparison to fixed-order NNLO pQCD calculations. We find that our data at both center-of-mass energies lie slightly under the NNLO pQCD predictions with recent PDF sets, which include the LHC data. Predictions using the earlier HERA PDF set fitted from DIS data align most closely to our data, compared to the other explored PDF sets. The kinematic limits explored in this analysis especially help to constrain gluon PDFs with parton momentum fraction $x>0.1$, where $x$ approximates the dimensionless jet $\xt{}= 2\pt/\sqrt{s}$.

The invariant cross sections as a function of jet $x_{T}$ were compared at $\sqrt{s} = 200$ and $510$~GeV. The ratios of the cross sections indicate scale-breaking effects similar to those observed at the Tevatron and LHC, but comparisons to NNLO pQCD calculations and \pythia{}~6 indicate consistent trends with our measurements.

Our data can be used as a baseline to study the properties of the quark-gluon plasma produced in heavy-ion collisions for example the jet quenching phenomenon through nuclear modification factors for inclusive jets at $\sqrt{s}=$~200~GeV from RHIC. They also serve as a fundamental test of the perturbative QCD framework used to interpret the multitude of spin-dependent measurements published from the STAR Collaboration. 

\begin{acknowledgments}
We are grateful to the FastNLO and NNLOJet groups specifically to Klaus Rabbertz, Jo\~{a}o Pires, Alexander Huss, and Daniel Britzger for providing the interpolation grid that was used in this paper to calculate the NNLO pQCD predictions. We thank the RHIC Operations Group and SCDF at BNL, the NERSC Center at LBNL, and the Open Science Grid consortium for providing resources and support.  This work was supported in part by the Office of Nuclear Physics within the U.S. DOE Office of Science, the U.S. National Science Foundation, National Natural Science Foundation of China, Chinese Academy of Science, the Ministry of Science and Technology of China and the Chinese Ministry of Education, NSTC Taipei, the National Research Foundation of Korea, Czech Science Foundation and Ministry of Education, Youth and Sports of the Czech Republic, Hungarian National Research, Development and Innovation Office, New National Excellency Programme of the Hungarian Ministry of Human Capacities, Department of Atomic Energy and Department of Science and Technology of the Government of India, the National Science Centre and WUT ID-UB of Poland, German Bundesministerium f\"ur Bildung, Wissenschaft, Forschung and Technologie (BMBF), Helmholtz Association, Ministry of Education, Culture, Sports, Science, and Technology (MEXT), Japan Society for the Promotion of Science (JSPS), and Agencia Nacional de Investigacion y Desarrollo de Chile (ANID), Chile.

\end{acknowledgments}
\section*{Data availability}
The data behind the plots is publicly available from HEPdata~\cite{hepdata.181422}. The broader datasets that were used are not publicly available, and subject to access according to the STAR Collaboration policy.
\bibliography{apssamp}

\end{document}